\documentclass[twocolumn,amsmath,amssymb,aps,pra,floatfix,nofootinbib,superscriptaddress]{revtex4}
\usepackage{graphicx,graphics,times,color}
\usepackage{bm}
\usepackage{longtable}

\def\be{\begin{equation}}
\def\ee{\end{equation}}
\def\ba{\begin{eqnarray}}
\def\ea{\end{eqnarray}}
\def\la{\langle}
\def\ra{\rangle}

\begin{document}
\title{Information Transferring Ability of the Different Phases of a finite XXZ Spin Chain}
\author{Abolfazl Bayat}
\affiliation{Department of Physics and Astronomy, University College
London, Gower St., London WC1E 6BT, UK}
\author{Sougato Bose}
\affiliation{Department of Physics and Astronomy, University College
London, Gower St., London WC1E 6BT, UK}

\begin{abstract}
We study the transmission of both classical or quantum information through all the phases of a finite XXZ spin chain. This characterizes the
merit of the different phases in terms of their ability to act as a quantum wire. As far as quantum information is concerned, we need only
consider the transmission of entanglement as the direct transmission of a quantum state is equivalent. The isotropic anti-ferromagnetic spin
chain is found to be the optimal point of the phase diagram for the transmission of quantum entanglement when one considers both the amount of
transmitted entanglement, as well as the velocity with which it is transmitted. But this optimal point in the phase diagram moves to the Neel
phase when decoherence or thermal fluctuations are taken to account. This chain may also be able to transfer classical information even when,
due to a large magnitude of the noise, quantum information is not transmitted at all. For a certain range of anisotropies of the model, a
curious feature is found in the flow of quantum information inside the chain, namely, a hopping mode of entanglement transfer which skips the
odd numbered sites. Our predictions will potentially be testable in several physical systems.
\end{abstract}

\date{\today}
\pacs{03.67.Hk, 03.65.-w, 03.67.-a, 03.65.Ud.} \maketitle
\section{Introduction} \label{sec1}
Recently, condensed matter many-body systems have been viewed in the
light of quantum information. For example, the entanglement inherent
in them has been investigated \cite{vlatko}. One can, however, ask a
different question: how does information placed on one part of a
many-body system {\em pass through} such a system? Aside its
fundamental interest, this question may lead to mechanisms for
moving information over small distances. The idea is to use a finite
many-body system such as a spin chain (a chain of perpetually
interacting stationary spins -- a one dimensional magnet) as a
data-bus \cite{bose}. Many-body dynamics transports information
placed on a spin at one end of the chain to the spin at its other
end with a certain efficiency. This is an ``all solid-state" bus
whose spins and interactions, except for those at its very ends, are
never controlled. Applications could be in moving information
between quantum registers or for moving classical bits in nano-scale
spintronics. This area, reviewed in \cite{bose-review}, has mainly
focussed on perfecting the information transfer (information transmission) by clever means
-- special couplings \cite{couplings}, encoding \cite{encode},
pulsing \cite{twamley}, memory \cite{memory} etc.

 An interesting question from a condensed matter angle is how the above process of
 information transmission varies with the ``phase" of the spin chain. By ``phase" we mean both the form of the spin-spin interactions and the
 relevant ground state resulting from that interaction. In this
 context, only one study has been performed, which involves spin-1 chains \cite{sanpera}. Additionally, gapless phases have been
 shown to be generically bad for a ``slow" information transmission process that can take place between two spins coupled weakly to a
 many-body system \cite{plenio1}. The same slow information transmission process between spins coupled weakly to an anti-ferromagnetic (AFM) chain has also
 been studied \cite{lorenzo}. However, there is no investigation yet of information transmission as a function of the phases of the simplest, namely the spin-1/2 chain, when
 all spins are coupled equally strongly so that information transmission is fast. Instead,
 a majority of the work has simply assumed a
  fully polarized (symmetry broken) ferromagnetic (FM) initial state
 of the spin chain
 \cite{bose-review}.
 Here we study the process of information transfer through all phases
of a $S=1/2$ XXZ Heisenberg-Ising chain which models a range of
realistic materials and, according to Ref. \cite{phase_diagram}, is
the most important paradigm in low-dimensional quantum magnetism.
Using finite chains (the case relevant for information transmission) and exact
diagonalization, we identify the point in the phase diagram which
provides the optimal data-bus in absence of any encoding,
engineering, control etc. Interestingly, this turns out to be the
``isotropic" AFM phase which is the most interesting phase
\cite{phase_diagram} of the XXZ model. Here the ground state has
complete SU(2) symmetry and contains significant ``quantum"
correlations or entanglement. This phase is, perhaps, also the most
common, as it appears in the ubiquitous Hubbard model at strong
repulsion and half filling. Additionally most solid-state spin
chains such as the famous KCuF$_3$ \cite{phase_diagram}, engineered
atomic-scale spin chains \cite{hirjib} and doped fullerine
Sc$@$C$_{82}$ chains \cite{yasuhiro} are naturally AFM.

This study is an example of non-equilibrium dynamics in many-body
systems, currently a topic of intense activity \cite{quench}. Our
 dynamics is induced by suddenly coupling a single
 spin (the one bearing the information) with one end of a finite
 spin chain. For a range of
 phases, certain spin correlation functions behave curiously during this dynamics so that the initial state
  of the added spin {\em hops} through the chain {\em skipping alternate sites}.
 Additionally, information transmission exhibits contrasting behavior in the FM and AFM parts of the so-called XY
 phase and has a sharp jump at the boundary of the XY and FM phases.

The structure of this paper is as follows: in section \ref{sec2} we show that entanglement distribution through an arbitrary channel is
equivalent to the process of transferring a quantum state through the channel. In section \ref{sec3} we introduce our model, i.e., $XXZ$
Hamiltonian, and in section \ref{sec4} we consider the entanglement distribution via whole phase diagram of a $XXZ$ chain. This is followed by
an explanation in section \ref{sec5}. In section \ref{sec6} we characterize the effect of the channel. In section \ref{sec7} and section
\ref{sec8}, the thermal fluctuations and interactions with a bath are investigated respectively. Classical communication through this system is
the subject of section \ref{sec9}, which is followed by considering the information flow ``inside the chain" in section \ref{sec10}. In section
\ref{sec11}, we give some potential physical realizations which might test our results, while we summarize our results in section \ref{sec12}.

\section{Equivalence of State Transferring and Teleportation Models of Information Transmission}\label{sec2}

 In order to transfer information from one place to another we have to transfer
a state (say the state of a spin) which encodes some information.
 In particular, when we are thinking about quantum information transmission,
 to quantify the quality of
transmission, we compute the fidelity between the sent and the received state. Since this fidelity is dependent on the initial state it is
preferable to take the average value of the fidelity over all possible equi-probable initial states. This average fidelity makes it possible to
compare transmission quality of different channels and different schemes of information transmission. To send quantum information from sender to
receiver one can think about two different strategies. In the first strategy, which is called ``quantum state transferring'', the quantum state
is sent through the channel directly. Because of the interaction between the channel and quantum state they become entangled and state
transferring is imperfect in the sense that the fidelity between the received state and the initial state is less than one. On the other hand,
instead of using state transferring one can use teleportation for sending quantum information. Teleportation is based on a shared entangled pair
between sender and receiver which plays the role of the resource \cite{teleportation}. In this second strategy of sending quantum information,
the sender generates a maximally entangled pair, keeps one part, and send the other one to the receiver through the channel. This shares an
entangled pair between both sides of the channel and teleportation between sender and receiver can be used for information transmission.
However, this fact that the entanglement of the shared pair is not maximal makes the teleportation imperfect. The importance of the second
strategy is that we just send one part of the singlet state through the channel and it is not necessary to study the effect of the channel on an
arbitrary state. What we show in this section is that the average fidelity in both strategies are the same. This was already shown in
\cite{tele_Horedecki} using a different technique for arbitrary dimensions of the Hilbert spaces and here we prove it again, just for qubits,
using a much simpler language.

Let's start with the state transferring. In this case quantum state goes through the channel.
An arbitrary quantum channel $\xi$ is completely determined by a set of Kraus operators $\{K_m\}$ such that the output of the channel is
\begin{equation}\label{Kraus}
    \rho^{ST}_{r}=\xi(\rho_s)=\sum_m K_m\rho_{s} K_m^\dagger, \ \ \   \sum_m K_m^\dagger K_m=I,
\end{equation}
where $\rho_{s}$ is the input state of the channel, $\rho_{r}$ is
the output state received by the receiver and $ST$ stands for
``State Transferring". Here we start from the most general form of
a qubit state
$|\psi_s\ra=\cos{\theta/2}|0\ra+e^{i\phi}\sin{\theta/2}|1\ra$ as
the input. After interacting the pure input state
$\rho_{s}=|\psi_s\ra\la\psi_s|$ with the channel the output state
$\rho^{ST}_r$ (given by Eq. (\ref{Kraus}))  is generally a mixed
state. Fidelity between the received and the sent state is easily
computed as $F^{ST}(\theta,\phi)=\la\psi_s|\rho^{ST}_r|\psi_s\ra$
which is dependent on input parameters $\theta$ and $\phi$. To get
an input independent quantity we average the fidelity over all
possible input states, i.e. the surface of the Bloch sphere, with
uniform weight. With a straight forward computation we end up with
\begin{equation}\label{Fav_ST}
    F^{ST}_{av}=\frac{1}{4\pi}\int{F^{ST}(\theta,\phi)\sin{\theta}d\theta d\phi}=\frac{1}{3}+\frac{1}{6}\sum_m|Tr(K_m)|^2,
\end{equation}
where $Tr(.)=Trace(.)$.

 Now, we try to use the teleportation strategy for sending quantum information. To achieve this strategy we prepare
a pair of singlet state
\begin{equation}\label{singlet}
    |\psi^-\ra=\frac{|01\ra-|10\ra}{\sqrt{2}}.
\end{equation}
Then we keep one part of the pair in the sender and send the other part through the channel $\xi$. Since the first part in the sender does not
interact with the channel the whole effect of the channel is explained by
\begin{equation}\label{Kraus2}
    \rho_{out}=I\otimes \xi (|\psi^-\ra \la \psi^-|)=\sum_m I\otimes K_m|\psi^-\ra \la \psi^-| I \otimes K_m^\dagger.
\end{equation}
Generally, the output state $\rho_{out}$ is not a maximally
entangled state so when it is used as the resource of the standard
teleportation scheme \cite{teleportation} it gives an imperfect
teleportation in the sense that the final achievable fidelity is
less than one. In \cite{bose-teleportation} it has been shown that
teleporting $\rho_s$ with using noisy resource $\rho_{out}$
generates the following state as the output of the teleportation.
\begin{equation}\label{noisy_tele}
    \rho^{TP}_{r}=\sum_{m=0}^3 Tr(\rho_{out}E_m) \sigma_m \rho_s \sigma_m,
\end{equation}
where $TP$ stands for ``Teleportation", $E_m=\sigma_m |\psi-\ra \la \psi-|\sigma_m$ and $\sigma_m$ are Pauli matrices ($\sigma_0=I, \sigma_{1,2,3}=\sigma_{x,y,z}$).
Similar to the first strategy, fidelity of the received and the sent state is defined as
$F^{TP}(\theta,\phi)=\la\psi_s|\rho^{TP}_r|\psi_s\ra$ and average fidelity for input states is easily computed over the surface of the Bloch sphere. The
average fidelity of teleportation scheme is
\begin{eqnarray}\label{Fav_TP}
    F^{TP}_{av}&=&\frac{1}{4\pi}\int{F^{TP}(\theta,\phi)\sin{(\theta)}d\theta d\phi}\cr
    &=&Tr(E_0\rho_{out})+\frac{1}{3}\sum_{m=1}^3 Tr(E_m\rho_{out})\cr
    &=&\frac{1+2 Tr(E_0\rho_{out})}{3}.
\end{eqnarray}
The parameter $Tr(E_0\rho_{out})=\la \psi^-|\rho_{out}|\psi^-\ra$
is called {\em singlet fraction} and as it is clear from Eq.
(\ref{Fav_TP}) that it completely captures the quality of the
transmission. It is also clear from Eq. (\ref{Fav_TP}) that
to have an average fidelity above 2/3, which is accessible to the classical teleportation,
singlet fraction should exceed 1/2.
Using the form of $\rho_{out}$ in Eq. (\ref{Kraus2})
and expanding the singlet as Eq. (\ref{singlet}) one gets
$Tr(E_0\rho_{out})=\frac{1}{4}\sum_m|Tr(K_m)|^2$. Substituting
this value in Eq. (\ref{Fav_TP}) shows that
$F^{TP}_{av}=F^{ST}_{av}$.

Getting identical average fidelity in both strategies is a very important result in quantum communication which shows the average effect of a
channel can be captured just by transferring one part of the singlet state through the channel and computing the singlet fraction. However,
sharing an entangled pair between sender and receiver has an advantage, namely that after a few transmissions the total (generally noisy)
entanglement can be converted by local actions \cite{BDSW} to nearly a pure singlet. This can be used to transmit any state near perfectly using
quantum teleportation. So, because of the importance of the amount of entanglement shared between the the sender and receiver, and its above
proven equivalence to the more straightforward transmission of quantum states, we mainly focus on the entanglement distribution through the
phase diagram of the $XXZ$ Hamiltonian.

\section{Introducing the Model}\label{sec3}
  We consider a spin chain as a channel for information transferring and we study
the property of each phase of the chain on the quality of information transmission.
We take one of the most well known models in condensed matter physics, namely $XXZ$ spin chain.
The Hamiltonian of the open XXZ chain of length $N_{ch}$ is
\begin{equation}\label{Hamiltonian_ch}
    H_{ch}=J\sum_{i=1}^{N_{ch}-1}\{\sigma_i^x\sigma_{i+1}^x+\sigma_i^y\sigma_{i+1}^y+\Delta
    \sigma_i^z\sigma_{i+1}^z\},
\end{equation}
with $J$ being a coupling constant, $\Delta$ being the anisotropy
and $\sigma_k^{x,y,z}$ being Pauli matrices for site $k$. This
Hamiltonian has a rich phase diagram.
 For $\Delta=1$ and $J<0$ this interaction is the
 FM Heisenberg chain widely discussed in the context of quantum communication
\cite{bose,bose-review,couplings}. More interesting regimes
exist for $J>0$ and different values of $\Delta$
\cite{phase_diagram}. $\Delta<-1$ is the FM phase with a
simple separable biased ground state with all spins aligned to the
same direction. $-1<\Delta\leq1$ is called $XY$ phase, which is a
gapless phase and consist of two different legs, ferromagnetic half
($-1<\Delta<0$) and anti ferromagnetic part ($0\leq\Delta\leq1$).
$1<\Delta$ is called {\em Neel phase}, where the spectrum is gapped
and we get nonzero staggered magnetization. In the limit $\Delta\gg
1$ it takes the form of Neel states ($|010101...01\rangle$).

\section{Information Transmission through Whole Phase Diagram of $XXZ$ Hamiltonian} \label{sec4}
  Though information transmission can be investigated either classically or quantum mechanically we will primarily
examine quantum information transmission and devote one section later to classical information transmission in the same systems. Of course, the
most natural setting would be sending the state of a single spin through the chain. However, because of previous "equivalency" discussions in
section (\ref{sec2}), we will examine the transmission
  of one part of a two spin maximally entangled state of the form (\ref{singlet})
while a spin chain channel (spins 1 to $N_{ch}$) is in its ground state of some Hamiltonian $H_{ch}$. At time $t=0$, the interaction of the
$0$th and the $1$st spin is suddenly switched on while $0'$ is kept isolated from the rest. The ensuing dynamics transports the initial state of
the $0$th spin through the chain to the $N_{ch}$th spin with some efficiency, so that after a while $0'$ will be entangled with $N_{ch}$. As the
singlet has the same representation in any bases, the above entanglement transfer already subsumes within it ``state transfer in arbitrary
basis" and is thus very general.

 The reader may naturally question how general the above physical setting (couplings etc.) of transferring entanglement through a spin chain channel is. Indeed one could
have taken weaker or stronger or different couplings at the sending and receiving ends. However, weaker couplings generally lead to ``slow"
transfer schemes which will be susceptible to decoherence. On the other hand, if we really do have stronger couplings or different couplings
available at our disposal, we could just use them for the whole chain for faster and potentially better transfer, rather than using those
special couplings only at the ends. So we think that the most natural question to investigate is to simply place a spin encoding the unknown
state to be transmitted at one end of the chain, and couple it with the same coupling as present in the rest of the chain (which, as we know
from the previous section, is equivalent to the type of entanglement transmission considered by us). In any case, without putting some
restrictions on the coupling model at the ends, there is too much freedom in the problem, and it may not be possible to give a precise answer to
the effectiveness of a phase to transfer quantum information. Moreover, also note that we are not considering the generation of entanglement
from inside the spin chain, which is an altogether different problem \cite{Hannu-Abol-quench}, but merely the {\em transmission} of entanglement
through the chain.

 Note that for our scheme we require the chain initially in a unique
ground state $|\psi_g\ra_{ch}$ and this may have to be selected out
by applying an arbitrarily small magnetic field (for odd $N_{ch}$
AFM chain and the FM chain). The interaction between the $0$th and
the $1$st spins (the interaction turned on at $t=0$) of the channel
is assumed to be of the same form and strength as the rest of the
interactions, namely
\begin{equation}\label{HI}
   H_I=J(\sigma_0^x\sigma_1^x+\sigma_0^y\sigma_1^y+\Delta\sigma_0^z\sigma_1^z),
\end{equation}

With the $0'0$ singlet, the total length of the system considered is
thus $N=N_{ch}+2$ with the initial state being
\begin{equation}\label{init0}
   |\psi(0)\ra=|\psi^-\ra_{0'0}\otimes|\psi_g\ra_{ch},
\end{equation}
and the total Hamiltonian being
\begin{equation}\label{Ht}
   H=I_{0'}\otimes(H_{ch}+H_I).
\end{equation}
So that $0'$ never
interacts with the rest. Also note that $H$ is simply a Hamiltonian
of a single spin chain $0...N_{ch}$ of length $N+1$. As the aim is
entanglement distribution, we are interested at the times that the
 entanglement between spins $0'$ and $N_{ch}$ peaks.  By turning on
the interaction between spin $0$ and spin $1$ of the channel the
initial state evolves to the state
$|\psi(t)\ra=e^{-iHt}|\psi(0)\ra$ and one can compute the density
matrix $\rho_{ij}=tr_{\hat{ij}}\{|\psi(t)\ra\la \psi(t)|\}$ where
the meaning of $tr_{\hat{ij}}$ is the trace over whole of the
system {\em except} sites $i$ and $j$ (We fix $i=0'$ in this
paper). The general form of a two spin density matrix $\rho_{ij}$
in $XXZ$ systems in the computational
($|00\rangle,|01\rangle,|10\rangle,|11\rangle$) basis are
\cite{wang-zanardi},
\begin{eqnarray}\label{general_form}
\rho_{ij}=\left(%
\begin{array}{cccc}
  u^+ & 0 & 0 & 0 \\
  0 & w^+ & z & 0 \\
  0 & z & w^- & 0 \\
  0 & 0 & 0 & u^- \\
\end{array}%
\right)
\end{eqnarray}
where all the elements of the matrix are real and they can be
written in terms of one and two point correlations,
\begin{eqnarray}\label{elements}
  u^{\pm} &=& \frac{1}{4}\{1+\la\sigma^z_i(t)\sigma^z_j(t)\ra\pm\la\sigma^z_i(t)\ra\pm\la\sigma^z_j(t)\ra\} \cr
  w^{\pm} &=& \frac{1}{4}\{1-\la\sigma^z_i(t)\sigma^z_j(t)\ra\mp\la\sigma^z_i(t)\ra\pm\la\sigma^z_j(t)\ra\} \cr
  z&=&
  \frac{1}{4}\{\la\sigma^x_i(t)\sigma^x_j(t)\ra+\la\sigma^y_i(t)\sigma^y_j(t)\ra\},
\end{eqnarray}
where $\sigma^\alpha_j(t)=e^{iHt}\sigma^\alpha_je^{-iHt}$ (for all
$\alpha=x,y,z$) is the Heisenberg picture of $\sigma^\alpha_j$ and
$<>$ means expectation value according to the initial state
(\ref{init0}). The concurrence as a measure of entanglement
\cite{concurrence} for this general density matrix
(\ref{general_form}) is $E=2 max (0, |z|-\sqrt{u^+u^-})$ which is
a function of time dependent correlators and expectation values.
Noteworthy, is that for $\Delta>-1$, when the initial state of the
channel is {\em not} symmetry broken, then symmetry considerations
and the fact that $0$ and $0'$ are initially anti-correlated in a
singlet, imply that the entanglement between $0'$ and $j$ can be
written as
\begin{equation}\label{ent_conc}
   E_{0'j}=max (0,
|\la\sigma^x_0(0)\sigma^x_{j}(t)\ra|-\frac{1}{2}\la\sigma^z_0(0)\sigma^z_{j}(t)\ra-\frac{1}{2}),
\end{equation}
which is solely written in terms of the two-time correlation functions of the spin
chain $0...N_{ch}$. It should be noticed that though two point
correlations of the $XXZ$ Hamiltonian have been studied
intensively in the literature and their asymptotic behavior is
known, the correlations here are different since they are
computed in terms of the initial state (\ref{init0}) which is not
the ground state of $H$. In addition, if one
ignores (traces out) spin $0'$, our study can be regarded as an
analysis of two time correlation functions during the
non-equilibrium dynamics that ensues when the interaction of a
spin in a random state with one end of a spin chain is switched
on. Singlet fraction of the state $\rho_{0'N}$, which was shown
that is directly related to the average fidelity of state
transferring, can be computed also from $\rho_{0'N}$ easily
\begin{equation}\label{Ft}
   F=\la\psi^-|\rho_{0'N}|\psi^-\ra=\frac{1}{2}(w^++w^--2z).
\end{equation}

\begin{figure} \centering
    \includegraphics[width=7cm,height=5cm,angle=0]{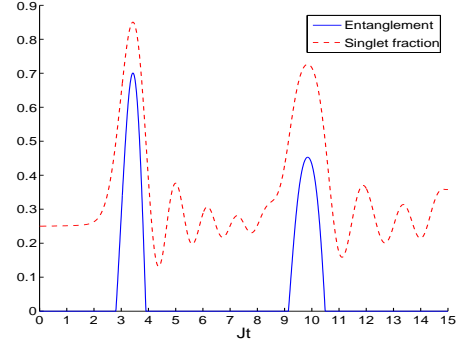}
    \caption{(Color online) Entanglement and singlet fraction in terms of time for a chain of length $N=20$ and $\Delta=1$.}
     \label{Fig1}
\end{figure}
In Fig. \ref{Fig1} we have plotted both entanglement and singlet
fraction of $\rho_{0'N}$ in terms of time for a particular point
in the phase diagram, namely $\Delta=1$. As it can be seen from
figure, singlet fraction always oscillates while the entanglement
just peaks at certain times which we call optimal time $t_{opt}$.
When entanglement peaks, singlet fraction also has a peak which
shows that final state is more similar to singlet than other Bell
states.
 \begin{figure} \centering
    \includegraphics[width=7cm,height=5cm,angle=0]{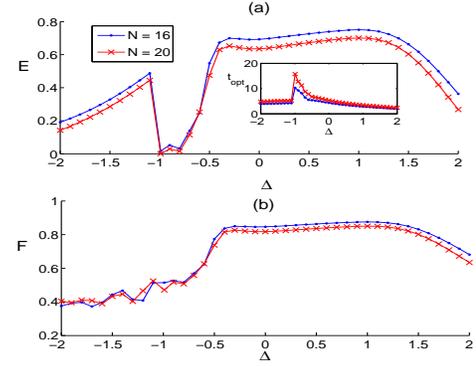}
    \caption{((Color online) a) Attainable entanglement in
the first peak in terms of $\Delta$ for different lengths (J=1). Inset shows
the optimal time $t_{opt}$ that the peak happens during the evolution. b) Singlet fraction $F$ at $t_{opt}$ in whole
phase diagram.}
     \label{Fig2}
\end{figure}
 The time that one can afford to wait for the entanglement
between $0'$ and $N_{ch}$ to attain a peak is restricted by
practical considerations such as the decoherence time, required
speed of connections in a quantum network etc. So we restrict
ourselves to the first peak of the entanglement in time. To compare
the performance of different phases of the Hamiltonian
(\ref{Hamiltonian_ch}) in transferring the entanglement, we have
plotted the amount of entanglement in its first peak ($t=t_{opt}$) in terms of
anisotropy $\Delta$ for the chains with different lengths in Fig.
\ref{Fig2}a and the associated singlet fraction at the same time in Fig. \ref{Fig2}b.
One interesting feature is that the entanglement
transmitted dips on the XY side of $\Delta=-1$ and sharply rises on
its FM side which captures the first order phase transition at this
point (that such a change in behavior is seen despite the finite
size is interesting). In addition, this transition is also marked by
a steep rise in the time required to reach the first peak in
entanglement. This is shown in the inset to Fig. \ref{Fig2}a,
which also shows that the speed at which entanglement is propagated
increases monotonically in the $XY$ regime as one goes from
$\Delta=-1$ to $1$. This fact is commensurate with the spin wave
velocity increasing with $\Delta$ as $\sin
(\cos^{-1}\Delta)/\cos^{-1}\Delta$ in this regime \cite{Saryer} which we will discuss it in more detail later.
In the XY phase we can recognize two distinct regimes. In its FM
sector ($-1<\Delta<0$) entanglement falls rapidly by decreasing
$\Delta$, while in its AFM sector ($0\leq\Delta\leq1$) the
entanglement is always good and increases by increasing $\Delta$.
After $\Delta=1$, when the transition from the $XY$ to the Neel
phase happens, the entanglement starts falling with increasing
$\Delta$, as the Ising term $\sigma^z_j\sigma^z_{j+1}$ dominates
which, by itself, does not transfer entanglement. Note also a subtle
feature that even outside the $XY$ regime, for $\Delta>1$,
entanglement falls much slower with $|\Delta|$ than for $\Delta<-1$.
In general, AFMs are thus better, even with similar degrees of
anisotropy. Fig. \ref{Fig2}a also shows that the isotropic AFM
Heisenberg interaction ($\Delta=1$) not only is the best for
transferring the highest amount of entanglement in the entire phase
diagram, but also it has the highest speed in the $XY$ phase.
In Fig. \ref{Fig2}b, where singlet fraction $F$ has been plotted in the entire of phase diagram,
in the $FM$ phase ($\Delta<-1$) singlet fraction is always less than 1/2 which shows despite the fact that entanglement is non-zero,
quantum communication has no benefit over classical communication. Same thing will happen
in the Neel phase when $F$ becomes less than 1/2 for quite large $\Delta$'s.
\begin{figure}
\centering
    \includegraphics[width=6cm,height=5cm,angle=0]{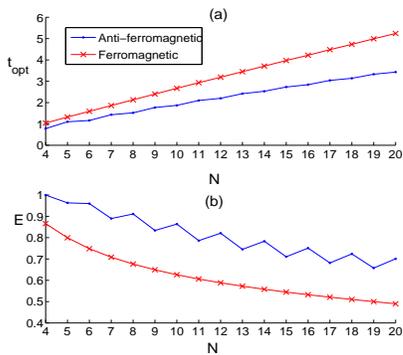}
    \caption{(Color online) a) Optimal time $t_{opt}$ for both FM ($J=-1$) and AFM ($J=+1$)
    in terms of length $N$. b) Entanglement at the first peak versus the
    length $N$ for both FM ($J=-1$) and AFM ($J=+1$) chains when $\Delta=1$.}
    \label{Fig3}
\end{figure}

 The effect of the length of the chain on the quality of transmission has
been shown in Fig. \ref{Fig3}. We only concentrate on the best
$\Delta=1$ (isotropic AFM) point, as it is the best point in the
phase diagram, and we compare the results with FM chains which have
been predominantly studied so far. In Fig. \ref{Fig3}a we
have plotted the time at which the first peak in entanglement for
different lengths. It is clear that the speed of entanglement
transmission through the AFM ($J>0$) chain is higher than FM ($J<0$)
chain independent of the length. In Fig. \ref{Fig3}b the
amount of entanglement in the first peak has been compared for both
the AFM and FM cases, from which it is clear the the entanglement
transmitted in the case of AFM chain has a distinctively higher
value irrespective of length. Note also a visible even-odd effect on
the amount of entanglement transmitted (and hence on two-time
correlations) which will be interesting to observe in finite chains.

 We have shown that in absence of any of the sophisticated techniques
for perfecting spin chain communications, which come at a price, and
may be hard to implement and also if one wanted to transfer
information fast (i.e, refrain from very weak couplings), the
isotropic AFM is the best channel in the entire phase diagram of the
XXZ chain. We now estimate the efficiency with which local
processing at the opposite ends of the spin chain and classical
communication between them (a process called entanglement
distillation \cite{BDSW}) can establish a {\em nearly perfect singlet} for an
isotropic AFM channel. For instance by using the recurrence algorithm for distillation \cite{BDSW} in a chain of length 10, for
which entanglement $E=0.8638$, starting from 9 impure pairs on
average leads to a nearly singlet state with entanglement $E=0.9920$
after 7 iterations, and for a chain of length 20, for which
entanglement $E=0.7162$, we need to start with 17 impure pairs to
get a singlet state with entanglement $E=0.9926$ after 9 iterations.
This perfect singlet can then be used for sending quantum states
perfectly through teleportation. It is worth pointing out that in
different phases the spin chains represent different types of
quantum channels. While in the FM phase, it is known to be an
amplitude damping channel (transmits $|0\rangle$ and $|1\rangle$
asymmetrically \cite{bose}), the $\Delta=1$ point affects a so
called depolarizing channel (also noted in \cite{lorenzo}) where
$\rho_{0'N_{ch}}$ is the mixture of the singlet state and the
Identity. Such $\rho_{0'N_{ch}}$ is particularly suited for
distillation protocols \cite{BDSW}.

\section{Explanation} \label{sec5}
When the phase of the system changes, not only the Hamiltonian causing the time evolution varies, but also the ground state, and consequently
the initial state (\ref{init0}) varies, and we have a different behavior for information transmission through the chain. Results of the Fig.
\ref{Fig2}a shows a dramatic and discontinuous change of entanglement at $\Delta=-1$ which is related to a first order phase transition at this
point and two completely different class of ground states of the $XY$ and the ferromagnetic phase. At point $\Delta=+1$ entanglement falls
continuously when we go from $XY$ phase to the Neel phase. This continuous change represents a second order phase transition at this point.
Beside these two phase transition points there is a sharp drop of entanglement around $\Delta=-0.5$ which is very peculiar since there is no
phase transition at this point. Also from the inset of Fig. \ref{Fig2}a it is clear that the optimal time which one has to wait to get a peak
goes up drastically for $\Delta<-0.5$. This strange property inside the $XY$ phase is certainly not because of a phase transition. The reason of
this slow dynamics and bad transmission is hidden behind an intrinsic property of the spin chain namely, the spin wave velocity.

Field theoretic techniques have been used to capture the
asymptotic behavior of correlation functions in spin chains \cite{fabricius}. For the general $XXZ$ Hamiltonian, it
fails to get all prefactors and
exact solutions, but it is able to get the qualitative behavior of
correlations successfully in the thermodynamic limit. Correlation
functions in our problem are different from those obtained by field theory
in at least two ways. First of all, we consider very finite chains,
since the idea of using spin chains as quantum channels is
valuable only for a finite distance. Secondly, all the dynamical correlations
which are computed asymptotically are associated with the ground
state of the system while in our problem correlations are computed
for the initial state (\ref{init0}) which is {\em not} the ground state.
Despite these differences, we still can use some well known results
of the field theoretic techniques. For example, the dynamical correlation
functions in the $XY$ phase ($-1<\Delta<1$) in the asymptotic thermodynamical limit have the following form \cite{fabricius}:
\begin{equation}\label{eta}
\langle \sigma^\alpha_j (0)\sigma^\alpha_k (t)\rangle \sim (-1)^{|j-k|}\frac{1}{(|j-k|^2-v_F^2 t^2)^{1/2 \eta_\alpha}},
\end{equation}
where $\alpha=x,y,z$ and
\begin{equation}
\eta_x=\eta_y=1/\eta_z=1-\frac{cos^{-1}\Delta}{\pi}.
\end{equation}
Moreover, $v_F$ in Eq. (\ref{eta}) is the spin-wave velocity
(this quantifies the propagation velocity of
excitations in the chain) which
has the following form
\begin{equation}\label{vF}
 v_F\propto \frac{\sin (\cos^{-1}\Delta)}{\cos^{-1}\Delta}.
\end{equation}
Unfortunately, the above asymptotic forms of $\langle \sigma^\alpha_j (0)\sigma^\alpha_k (t)\rangle$, valid for $|j-k|>>v_Ft$, are {\em
singular} specifically at $|j-k|\sim v_F t$, which is the regime relevant to optimal quantum communication from $j$th to the $k$th site (i.e.,
when the information, possibly traveling at a velocity $v_F$ reaches its destination). So one can only use some aspects reliably from the above
formulae. One of this is the velocity $v_F$ of propagation of the correlations (and hence information).
\begin{figure}
\centering
    \includegraphics[width=6cm,height=5cm,angle=0]{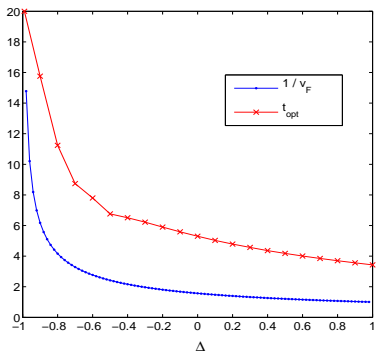}
    \caption{(Color online) $1/v_F$ (for infinite chain) and $t_{opt}$ (for a chain of length $N=20$) versus $\Delta$.
     }
    \label{Spin_wave}
\end{figure}
In Fig. \ref{Spin_wave} we have plotted $1/v_F$ (which is for an infinite chain) and $t_{opt}$ (for a chain of length $N=20$) in terms of
$\Delta$. As Fig. \ref{Spin_wave} clearly shows, both of these quantities behave in a {\em strikingly similar} way (the gap between two curves
is not important since one can multiply them by some constants). As a consequence, for $\Delta<-0.5$ the propagation velocity is very slow, and
one has to wait a long time to receive some information at the other side of the chain. This very slow dynamics also means that we will get a
sharp fall in the entanglement as well as all other quantities which propagate through the chain unless we are willing to wait for very very
long times.

As far as the question of why the isotropic ($\Delta=1$ point) is the best point in the phase diagram in terms of a maximum of entanglement,
recall that entanglement is given in terms of some dynamical correlation functions as in Eq. (\ref{ent_conc}). Remember though, that these
correlations are {\em not} evaluated for the ground state, so cannot be strictly substituted by the known dynamical correlation functions to get
any quantitative information, but perhaps only a qualitative picture, as we discuss. One can see from Eq.(\ref{eta}) that the $\Delta=1$
 point is the best for the propagation of correlations along the $z$ direction. At points with
$\Delta<1$ the $z$ component of correlations does not propagate as well as the $x$ component, and, in fact, near to $\Delta=-1$ it is expected
not to propagate at all ($\eta_z \approx \infty$). Thus the term $-\frac{1}{2}\la\sigma^z_0(0)\sigma^z_{j}(t)\ra$ in Eq. (\ref{ent_conc}) for
entanglement, which is positive, contributes more and more as we approach $\Delta=1$ and gives a higher entanglement. It is true that as we
approach the isotropic point from $\Delta<1$ side, the $\eta_x$ rises (i.e., propagation of correlations in the $x$ direction deteriorates
somewhat). However, it must be that the gain from the better propagation of correlations in the $z$ direction more than compensates for the
deterioration of the propagation of correlations along the $x$ direction. The reason is that $\eta_z$ changes from $\infty$ to $1$ (huge gain),
while $\eta_x$ only goes from $0$ to $1$.

As far as the intriguing dip after $\Delta=-0.5$ is concerned, we are not yet in a position to explain it. It seems that the behavior expected at $\Delta \rightarrow -1$ where
both the velocity of correlations and their propagation quality along the $z$ direction are worst, starts to happen quite a bit {\em before} the actual point.

\section{Channel Characterization and Even-Odd Effect} \label{sec6}
As it is clear from Fig. \ref{Fig3}b in the case of
AFM chain, entanglement has a zig-zag behavior when
$N$ varies while it behaves uniformly for FM chains.
This even-odd effect for AFM chains has a
fundamental reason. In even chains when $\Delta>-1$ the total
magnetization of the ground state is always zero and because of
the rotational symmetry in the ground state one can exchange all
$|0\ra$'s and $|1\ra$'s while the ground state remains unchanged.
In other words, in even chains for $\Delta>-1$ we always have
\begin{equation}\label{sym}
  \sigma_x^{\otimes N_{ch}}|\psi_g\ra_{ch}=|\psi_g\ra_{ch}.
\end{equation}
This symmetry, which is absent in FM chains and also in each of the doubly degenerate ground
states of the odd chains, has a profound effect on the
transmission characteristics of the chain. In even chains, the
effect of the chain is completely recognized by the Kraus
operators $\{ \sqrt{p_I}I, \sqrt{p_x}\sigma_x, \sqrt{p_y}\sigma_y,
\sqrt{p_z}\sigma_z\}$. Where, $p_{I,x,y,z}$ are positive and their
summation is equal to 1 so, one can explain the effect of this
channel such that it applies one of the Pauli operators (including
identity) with some probability to the input state. Obviously,
these probabilities are dependent on the length $N$, time $t$ and
anisotropy $\Delta$. So using the above Kraus operators we get the
following form for the state $\rho_{0'N_{ch}}$
\begin{eqnarray}\label{rho_final}
  \rho_{0'N_{ch}}&=&p_I(t)|\psi^-\ra \la\psi^-|+p_x(t)|\phi^-\ra \la\phi^-|\cr
  &+&p_y(t)|\phi^+\ra \la\phi^+|+p_z(t)|\psi^+\ra \la\psi^+|,
\end{eqnarray}
where,
\begin{eqnarray}\label{Bellstates}
  |\psi^{\pm}\ra&=&\frac{|01\ra\pm |10\ra}{\sqrt{2}},\cr
 |\phi^{\pm}\ra&=&\frac{|00\ra\pm |11\ra}{\sqrt{2}},
\end{eqnarray}
are Bell states. Thus it means that $\rho_{0'N_{ch}}$ is diagonalaized in the Bell basis.
This channel is called {\em Pauli channel} in the
literature. Since in the Hamiltonian (\ref{Ht})
there is no difference between $x$ and $y$ directions we always have $p_x=p_y$ for $XXZ$ chain. At the point $\Delta=1$
where all directions become identical we have $p_x=p_y=p_z$ and channel is the famous depolarizing channel.

For the case of odd $N$, characterization of the channel is not yet known and one can just consider it numerically.
Since in odd chains the ground state of the system is degenerate, to take one of them we apply a small magnetic field in the $z$ direction
to break the symmetry. In this case the total magnetization is $\pm 1$ (dependent on the direction of the magnetic field: $\pm z$) and
the symmetry (\ref{sym}) does not hold anymore.

For $\Delta<-1$ since the ground state is ferromagnetic and all
spins are aligned the type of the channel is amplitude damping
\cite{bose} so then the even-odd effect vanishes and channel
behaves uniformly for all $N$. It worths to mention that in entanglement distillation procedures \cite{BDSW}
Werner states, which are a mixture of Bell states, are distilled more easily than the other states \cite{BDSW} so
transferring the singlets through a Pauli channel has this advantage that the final state is very close
to a Werner state (at $\Delta=1$ it is exactly a Werner state) and one can distill them more easily than those
which are gained through the transmission of other channels such as amplitude damping.

\begin{figure}
\centering
    \includegraphics[width=6cm,height=5cm,angle=0]{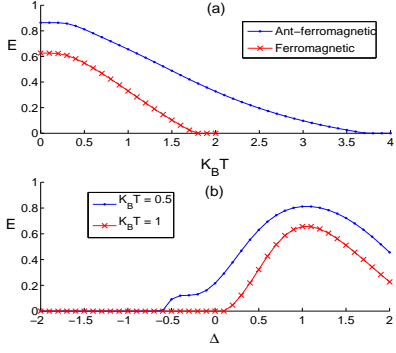}
    \caption{(Color online) a) Entanglement in terms of temperature in a chain of length $N=10$ for the isotropic case ($\Delta=1$) in both FM ($J=-1$)    and AFM ($J=+1$) phase.
    b) Entanglement in whole phase diagram for different temperatures in the chain of length $N=10$.}
    \label{Fig4}
\end{figure}

\section{Thermal Fluctuations} \label{sec7}
  Generally, when system is in non zero temperature, the state of
the channel before evolution is described by a thermal state
$\frac{e^{-\beta H_{ch}}}{Z}$ instead of the ground state, where
$\beta=1/K_BT$ and $Z$ is the partition function. So in
this case the initial state of the system is

\begin{equation}\label{rho0}
  \rho(0)=|\psi^-\ra
\la\psi^-|\otimes \frac{e^{-\beta H_{ch}}}{Z}.
\end{equation}
We assume that the thermalization time-scale of the system is large so that
one can consider the unitary dynamics starting the initial state (\ref{rho0}).
So, after time $t$
system evolves to $\rho(t)=U\rho(0)U^\dagger$ and the target state $\rho_{0'N_{ch}}(t)$ can be gained as before by tracing out the bulk of the chain
$\rho_{0'N_{ch}}(t)=tr_{\hat{0'N_{ch}}}\{\rho(t)\}$.
Entanglement of the state $\rho_{0'N_{ch}}(t)$ at its optimal time has been plotted in Fig. \ref{Fig4}a in terms of
initial temperature for both FM ($J=-1$) and AFM chains ($J=+1$).
As it is clear from the figure, increasing the temperature always destroys the entanglement but it has less effect
on AFM chain.

In Fig. \ref{Fig4}b the entanglement in whole phase diagram has
been plotted for different temperatures. When temperature rises
entanglement survives more for fully symmetric Heisenberg point
($\Delta=1$). Specially FM phase is highly sensitive to
thermal fluctuations, and entanglement is destroyed rapidly when
temperature rises. Furthermore, we found that, optimal time
$t_{opt}$ which the entanglement peaks is almost independent of
the temperature and varies very slowly in the entire of phase
diagram.

\begin{figure}
\centering
    \includegraphics[width=6cm,height=5cm,angle=0]{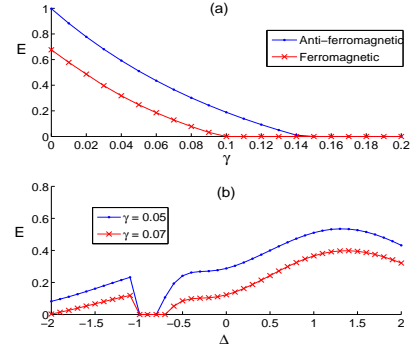}
    \caption{(Color online) a) Entanglement in terms $\gamma$ in a chain of length $N=8$ for the isotropic case ($\Delta=1$) in both FM ($J=-1$)
    and AFM ($J=+1$) phase.
    b) Entanglement in whole phase diagram for different noise strength $\gamma$ in the chain of length $N=8$.}
    \label{Fig5}
\end{figure}

\section{Interaction with Bath and Decoherence Effect} \label{sec8}
 In practical situations it is impossible to isolate a quantum
system from its environment. In the case of Markovian interaction
between the system and the environment, a Lindblad equation
describes the evolution of the system
\begin{equation}\label{Noiseevol}
 \dot{\rho}=-i[H,\rho]+\ell (\rho),
\end{equation}
where $\ell(\rho)$ is the Markovian evolution of the state
$\rho$. Let us assume an environment which has no preferred
direction. Eventually the interaction should have the following
form
\begin{eqnarray}\label{Leinblad}
    \ell(\rho)=-\frac{\gamma}{3}
    \sum_i\sum_{\alpha}\{\rho-\sigma^{\alpha}_{i}\rho\sigma^{\alpha}_
    {i}\},
\end{eqnarray}
where index $i$  takes $0',0,...,N_{ch}$ and $\alpha$ gets
$x,y,z$, and the coefficient $\gamma$ stands for the rate of
decoherence. In Fig. \ref{Fig5}a we have plotted the entanglement
in terms of noise strength $\gamma$ for both FM
($J=-1$) and AFM ($J=+1$) chain. Since the figure
clearly shows entanglement decays exponentially by increasing
$\gamma$ but like the thermal effect AFM chain is
more resistive against Markovian noise. In Fig. \ref{Fig5}b, we
have plotted the attained entanglement in when $\Delta$ varies in
whole phase diagram. As the figure shows the noise effect always
kills the entanglement and similar to the thermal fluctuations
optimal time is almost independent of noise parameter $\gamma$.
Surprisingly the best point in the phase diagram is not the case
of $\Delta=1$ and it moves down inside the Neel phase. The reason
of this interesting phenomena comes from the velocity of dynamics.
Since the evolution in the Neel phase is faster (see the inset of Fig. \ref{Fig2}a) and entanglement
peaks earlier, decoherence has less opportunity to interfere and
shows its destructive effect.

\section{Classical Communication} \label{sec9}
 We now comment on the classical information transmission
through spin chains, which may be interesting for spintronics. To
quantify the amount of classical information which each channel
can transmit concept of the classical capacity has been
introduced. The classical capacity of the channel $\xi$ (introduced by
the Kraus operators (\ref{Kraus}) in a very general case) gives the
maximum amount of classical information that can be reliably
transmitted per channel use. In calculating the classical capacity
it is necessary to perform a maximization over multiple uses of
the channel
\begin{equation}\label{Capacity}
 C=max_n  \frac{C_n}{n},
\end{equation}
where $C_n$ is the classical capacity of the channel $\xi$ which
can be achieved if the sender is allowed to encode the
information on codewords which are entangled only up
to $n$-parallel channel uses. The value of $C_n$ is obtained
by maximizing the Holevo information \cite{Holevo}
at the output
of $n$ parallel channel uses, over all possible input ensembles $\{p_i,\rho_i\}$, i.e.
\begin{equation}\label{Cn}
 C_n=max_{\{p_i,\rho_i\}}  H_n(\xi^{\otimes n},\{p_i,\rho_i\}),
\end{equation}
where $H_n(\xi^{\otimes n},\{p_i,\rho_i\})$ is the Holevo information which is defined as
\begin{equation}\label{Holevo_bound}
 H_n=\{S(\xi^{\otimes n}(\sum_i p_i\rho_i))-\sum_i p_i \xi^{\otimes n} (\rho_i)\}.
\end{equation}
Here $p_i$'s are probabilities, $\rho_i$'s are $n$-qubit codewords
(either entangled or separable) and $S$ is the von Neumann entropy.
Unfortunately computing the classical capacity is an extremely hard
task since it needs a very difficult maximization. But recently the
classical capacity of the depolarizing channel has been computed
\cite{depolarizing_capacity} and it was shown that this capacity
can be achieved by encoding messages as products of pure states
belonging to an orthogonal basis, and using measurements which are
products of projections onto this same orthogonal basis. So due to
the fact that entanglement does not increase the capacity of the
depolarizing channel all maximization shrinks to compute the
single shot capacity $C_1$ as the real capacity of the channel.

As we discussed in section (\ref{sec4}), for even chains $XXZ$
Hamiltonian is a Pauli channel. At isotropic point ($\Delta=1$) it
is a depolarizing channel, which $C_1$ is the real capacity and
entangled inputs do not increase it. This motivates us to study
{\em single-shot} classical capacity of the $XXZ$ Hamiltonian for {\em
pure orthogonal} input states. However the single-shot capacity
which is computed over pure orthogonal input states is not
necessarily the real capacity of the channel (except at the point
$\Delta=1$) but at least it gives us a lower bound of the
classical capacity. To have the form of Pauli channel, we also
restrict our study just to the even chains.

We start with the most general form
of the orthogonal pure qubit states
\begin{eqnarray}\label{psi1-psi2}
    |\psi_1\ra&=&\cos{\frac{\theta}{2}}|0\ra+e^{i\phi} \sin{\frac{\theta}{2}}|1\ra \cr
    |\psi_2\ra&=&\sin{\frac{\theta}{2}}|0\ra-e^{i\phi} \cos{\frac{\theta}{2}}|1\ra,
\end{eqnarray}
where $0\leq \theta\leq\pi$ and $0\leq \phi\leq 2\pi$. For input ensemble,
we associate the probability $p_1$ to the input state $\rho_1=|\psi_1\ra \la\psi_1|$ and similarly
probability $p_2$ to the state $\rho_2=|\psi_2\ra \la\psi_2|$. When each of these states goes through the channel
we get
\begin{equation}\label{rhoF}
 \xi(\rho_i)= p_I \rho_i+p_x \sigma_x \rho_i \sigma_x+p_y \sigma_y \rho_i \sigma_y+p_z \sigma_z \rho_i \sigma_z, \ \ i=1,2.
\end{equation}
Where in the above equation $p_{I,x,y,z}$ are dependent on time
$t$ and anisotropy $\Delta$. It is easy to see that
$S(\xi(\rho_1))$ and $S(\xi(\rho_2))$ are equal and independent of
$\phi$. Thus the second term in the Holevo information
(\ref{Holevo_bound}) is
$p_1S(\xi(\rho_1))+p_2S(\xi(\rho_2))=S(\xi(\rho_1))$, which is
independent of $p_1$ and $p_2$ and it is just dependent on
$\theta$. We can easily maximize the first term in the Holevo
information (\ref{Holevo_bound}) for all values of $\theta$ by
choosing $p_1=p_2=1/2$ such that $S(\xi(p_1\rho_1+p_2\rho_2))=1$
so to maximize the Holevo information $H_1$ one should just find
$\theta=\theta_{opt}$ such that minimize $S(\xi(\rho_1))$.

Our anaytic computation shows that
\begin{eqnarray}\label{theta_opt}
    if \ \ p_z>p_x: \ \ \theta_{opt}&=&0 \ \ or \ \ \pi . \cr
    if \ \ p_z<p_x: \ \ \theta_{opt}&=&\pi/2.\cr
    if \ \ p_z=p_x: \ \ \theta_{opt}&=&arbitrary,
\end{eqnarray}
where the situation $p_z=p_x$ is associated to the depolarizing
channel ($\Delta=1$) which for any value of $0\leq \theta\leq\pi$
the classical capacity is achieved. Important point is that the
optimal input ensemble is independent of phase $\phi$ which gives
us a lot of degrees of freedom for input states. This also was
expected due the symmetry of the $x$ and $y$ directions in our
Hamiltonian (\ref{Ht}).

\begin{figure}
\centering
    \includegraphics[width=6cm,height=5cm,angle=0]{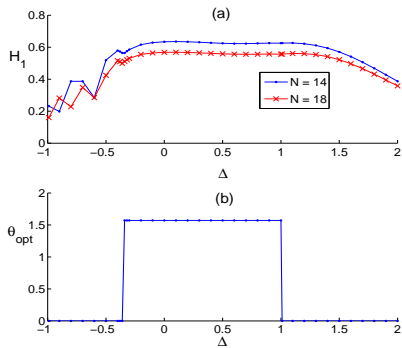}
    \caption{(Color online) a) Holevo information $H_1$ in terms of $\Delta$ for different lengths.
    b) Optimal angel $\theta_{opt}$ for input set of states in terms of $\Delta$.}
     \label{Fig6}
\end{figure}
In Fig. \ref{Fig6}a, we have plotted the classical capacity in terms of $\Delta$. This figure clearly shows that $H_1$ is quite flat in the $XY$
phase and it suddenly falls for $\Delta<-0.5$. {\em More interesting} result has been shown in Fig. \ref{Fig6}b where optimal $\theta$ has been
plotted in whole phase diagram. It shows that when we cross the point $\Delta=1$ from $XY$ phase to the Neel phase suddenly the optimal ensemble
changes from orthogonal states on the equator ($\theta=\pi/2$) of the Bloch sphere to the states on the poles ($\theta=0$). Within the $XY$
phase, around $\Delta=-0.35$ the optimal input ensemble changes such that optimal states for $-1<\Delta<-0.35$ are gained by $\theta=0$ and for
$-0.35<\Delta<1$ are obtained by $\theta=\pi/2$. In the the FM phase ($\Delta<-1$) the transmission is completely different and it is explained
as an amplitude damping channel \cite{bose}. For this channel it was shown that $C_1$ is achieved by the inputs given as Eq. (\ref{psi1-psi2})
for $\theta=\pi/2$ \cite{giovannetti-capacity}.

It is interesting to check the classical capacity of the channel
when it can not transmit quantum information. So for a chain of
length $N=8$ with the noise parameter $\gamma=0.3$, entanglement
can not be transferred because of the large noise (quantum
information transmission is impossible) but at optimal times for
isotropic case ($\Delta=1$) one gains $C_1=0.3931$ for
AFM chain ($J=+1$) and $C_1=0.1453$ for
FM chain ($J=-1$).

\begin{figure}
\centering
    \includegraphics[width=6cm,height=5cm,angle=0]{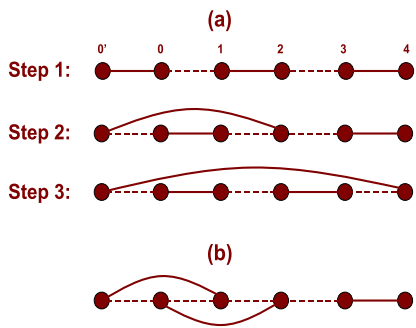}
    \caption{(Color online) a) Entanglement between site $0'$ and other sites in the chain during the
     evolution in an AFM chain ($\Delta=1$) of length $N=6$. b) One configuration of the states which are not accessible energetically.}
     \label{Fig_ver_5}
\end{figure}
\section{Entanglement Propagation through the Chain} \label{sec10}
 A curious feature emerges in the propagation of entanglement through
 chains with even numbers of spins. For $\Delta\geq 0$, there is never
any entanglement at any time between site $0'$ and odd sites and entanglement seems to {\em hop} through the chain. If one takes an approach
whereby one draws a bond for the presence of strong entanglement and dotted for very weak entanglement ($<0.1$), the open ended ground state
will be {\em depicted} as a dimer (remember it is not an exact dimer) \cite{wang}. Appending a singlet of spins $0$ and $0'$ at one end of the
chain, makes the total system look like a series of strongly entangled pairs next to each other (with weaker links between) and this is shown
for the $N=6$ case in step 1 of Fig. \ref{Fig_ver_5}(a). When the system evolves, the state of the system takes the form of step 2 in Fig.
\ref{Fig_ver_5}(a) and after a while it goes to the form of step 3 in Fig. \ref{Fig_ver_5}(a). To explain this curious effect, without losing
the generality, we consider the isotropic AFM. Clearly Fig. \ref{Fig_ver_5}(b), where a singlet between $0'$ and an odd site breaks 3 strong
bonds, is energetically not favored in course of a unitary dynamics starting as step 1 of Fig. \ref{Fig_ver_5}(a). So despite a finite (but
small) overlap between the state shown in Fig. \ref{Fig_ver_5}(b) and those in Fig. \ref{Fig_ver_5}(a) this state does not emerge through the
dynamics in the sense that its overlap with the state $|\psi(t)\ra$ never become higher than a certain value. Quantitatively, all moments of
Hamiltonian are conserved \cite{plenio1} during the evolution:
    $\forall n,\ \  \la
    H^n\ra=\la\psi(t)|H^n|\psi(t)\ra=\la\psi(0)|H^n|\psi(0)\ra$,
so energy ($E=\la H\ra$) and its variance ($\eta=\sqrt{\la H^2\ra-\la H\ra^2}$) are constant during the evolution. It means that only states
with energy expectation $\bar{E}$ for which $E-\eta<\bar{E}<E+\eta$, such as in Fig. \ref{Fig_ver_5}(a) can contribute in evolution, while those
as in Fig. \ref{Fig_ver_5}(b) cannot play a role. Note also that this curious phenomena, when recast in terms of two-time correlations, states
that $<\sigma_0^z(0)\sigma_j^z(t)>$ should be less than $-1/3$ only for the even sites $j$. Thus is potential physical systems where such
dynamical correlations is measurable, the hopping mode of transfer should be testable.

\section{Potential Physical Realizations} \label{sec11}
  We now mention some systems in which our results can be potentially tested, though there is some way to go for some of these systems,
  as local addressing of the spin to be suddenly coupled to the chain may be required. Recently there has been extensive
interest in finite spin chains such as fabricated AFM nano-chains \cite{hirjib}, and especially even-odd effects in such systems \cite{lounis}.
This can be one potential system where recently developed sensitive magnetometers \cite{lukin} can perhaps be used for verifying the
correlations and hence entanglement. Perhaps an STM tip encoding the spin to be transmitted can be brought close to one end of a finite array.
Finite chains of doped fullerines in nanotubes \cite{Briggs-chain} (such as AFM Sc$@$C$_{82}$ \cite{yasuhiro}) is the other alternative for
developing this idea. Spins in such systems have already been measured, and perhaps local electrical gates can give local control to couple in
the input qubit \cite{Briggs-chain}. Optical superlattices with atoms can realize an ensemble of finite spin chains \cite{Bloch} as well as the
switching on their interactions \cite{duan}. Barrier heights at regular intervals may be raised to create arrays of small lattice segments
(cells) of sizes $2$ and $N_{ch}$ with the repeating pattern $2,N_{ch},2,N_{ch},....$. The $0'0$ singlet and the finite chain ground states can
be created in the cells of sizes $2$ and $N_{ch}$ respectively as ground states (in fact, the former has already been accomplished
\cite{Bloch}). Next, again through global methods, the barriers between the $2$ site cells and the $N_{ch}$ site cells to their right have to be
lowered (simultaneously the barrier between the two sites of the cell of length $2$ has to be raised), so as to form superlattices with cells of
size $N_{ch}+2$ each. The subsequent dynamics will then exactly be as we have predicted and can potentially be verified through global time of
flight correlation measurements \cite{Bloch}. One can use ion traps where small spin systems are being realized \cite{Porras-Nature-2008}, as
well as implementing
 spin chains with trapped electrons \cite{Marzoli-Tombesi} where initializing individual spins and controlling
 the interaction at one end are both simple. NMR is another fruitful
 avenue for testing communication through spin chains \cite{NMR}.

\section{Summary} \label{sec12}
We have studied the transmission of both classical and quantum information through the all phases of the $XXZ$ Hamiltonian. This quantifies the
ability of each phase for information transmission. We found that in the absence of noise and thermal fluctuation isotropic Heisenberg
Hamiltonian ($\Delta=1$) is the best point of the phase diagram for information transmission, both in terms of its amount, as well as its speed.
The speed of propagation of the information, despite our finite open-ended case, fits strikingly well with the spin wave velocities known from
continuum limit field theoretic studies of the XXZ spin chain. When decoherence and thermal fluctuations are taken to account, the best point of
the phase diagram moves to the Neel phase, which due to a faster evolution, is less sensitive to these sources of noise. Furthermore, we showed
that the transmission through an even chain is characterized by the Pauli channel which has benefits in terms of immediate applicability of
entanglement distillation. We also studied the transmission of classical information through this channel. Optimal states for single-shot
classical capacity were identified and we realized that even when system is so noisy, such that quantum information is completely destroyed,
some classical information can be transferred. Studying the entanglement propagation through the chain showed that entanglement skips odd
numbered sites and manifests as a curious behavior of two time correlation functions during the non-equilibrium dynamics. It remains an open
problem to explain well the mysterious behavior of the dynamics which entanglement suddenly drops around the point $\Delta=-0.5$.

\section{Acknowledgement}
  SB is supported by an Advanced
Research Fellowship from EPSRC, through which AB is funded. SB is also supported by the QIP IRC,
Royal Society and Wolfson foundation.

\end{document}